\pdfoutput=1

\documentclass[11pt]{article}

\usepackage{ACL2023}

\usepackage{times}
\usepackage{latexsym}
\usepackage{svg}
\usepackage{amsmath}

\usepackage[T1]{fontenc}

\usepackage[utf8]{inputenc}

\usepackage{microtype}

\usepackage{inconsolata}

\title{Building Accurate Low Latency ASR for Streaming Voice Search}

\author{Abhinav Goyal, Nikesh Garera \\
  Flipkart \\
  \texttt{\{abhinav.goyal,nikesh.garera\}@flipkart.com}}

\begin{document}
\maketitle
\begin{abstract}
Automatic Speech Recognition (ASR) plays a crucial role in voice-based applications. For applications requiring real-time feedback like Voice Search, streaming capability becomes vital. While LSTM/RNN and CTC based ASR systems are commonly employed for low-latency streaming applications, they often exhibit lower accuracy compared to state-of-the-art models due to a lack of future audio frames. In this work, we focus on developing accurate LSTM, attention, and CTC based streaming ASR models for large-scale Hinglish (a blend of Hindi and English) Voice Search. We investigate various modifications in vanilla LSTM training which enhance the system's accuracy while preserving its streaming capabilities. We also address the critical requirement of end-of-speech (EOS) detection in streaming applications. We present a simple training and inference strategy for end-to-end CTC models that enables joint ASR and EOS detection. The evaluation of our model on Flipkart's Voice Search, which handles substantial traffic of approximately 6 million queries per day, demonstrates significant performance gains over the vanilla LSTM-CTC model. Our model achieves a word error rate (WER) of 3.69\% without EOS and 4.78\% with EOS while also reducing the search latency by approximately $\sim$1300 ms (equivalent to 46.64\% reduction) when compared to an independent voice activity detection (VAD) model.
\end{abstract}

\section{Introduction}

As an e-commerce platform in India, we need to cater to a variety of user bases, and a big part of that consists of users who cannot or do not want to type while interacting with the app, e.g., while searching for a product. For such users, interaction via a voice-based interface becomes an essential feature requiring an accurate and efficient Automatic Speech Recognition (ASR) system.

Recent years have witnessed the popularity of end-to-end ASR models, which have achieved state-of-the-art results \cite{e2e_asr_survey}. These models offer simplified training and inference processes and have demonstrated higher accuracy compared to traditional pipelines with separate acoustic, pronunciation, and language models. Common approaches for end-to-end ASR models include CTC (Connectionist Temporal Classification), AED (Attention-based Encoder-Decoder), and RNNT (RNN-Transducer) \cite{ctc, las, rnnt}.

However, streaming capability plays a pivotal role in choosing the most suitable ASR model. While non-streaming models can leverage the entire audio for text inference, streaming models have access only to past context, which can result in reduced accuracy. Nevertheless, streaming models provide immediate feedback, a critical requirement for consumer-facing applications like Voice Search. Additionally, low inference latency is essential to ensure a user-friendly experience, as delayed feedback can adversely impact usability.

Another challenge in streaming ASR applications is accurately detecting the end of speech (EOS). Conventional methods rely on standalone Voice Activity Detection (VAD) models, which operate independently from the ASR system and may not offer optimal accuracy.

In this work, we focus on developing a streaming ASR system for large-scale Hinglish Voice Search. Our objective is to enhance accuracy and reduce latency while preserving streaming capabilities. Specifically, we propose modifications to an LSTM and CTC based ASR system, aiming to bridge the gap between streaming and non-streaming ASR models. We also present a simple training and inference strategy that enables joint ASR and EOS detection within end-to-end CTC models, effectively reducing user-perceived latency in voice search. The contributions of this research can be summarized as follows:
\begin{itemize}
    \item Development of an accurate and efficient streaming ASR model based on LSTM, MHA (Multi-Head Attention), and CTC for Hinglish Voice Search;
    \item Introduction of a straightforward training and inference strategy to enable joint ASR and EOS detection within end-to-end CTC models, addressing the need for accurate EOS detection in streaming applications.
    \item Analysis of the impact of model modifications on reducing the performance gap between streaming and non-streaming ASR models.
\end{itemize}

Next, we discuss some related work in Section~\ref{sec:literature}. Section~\ref{sec:method} describe the model architecture we use, EOS integration and the inference method. We talk about the dataset and experimental setup in Section~\ref{sec:setup}. Finally, we conclude with a discussion on results and limitations in Section~\ref{sec:results}.

\section{Related Work}
\label{sec:literature}

CTC, the first E2E approach developed for ASR \cite{ctc}, has been widely used over the last few years \cite{ctc1,ctc2}. Although it provides simplicity, it makes a conditional independence assumption, that output token at any time doesn't depend on past tokens, which can make it sub-optimal. AED and RNNT models relax this assumption by leveraging past output tokens. While AED models like LAS (Listen, Attend and Spell) \cite{las} work very well for non-streaming tasks, they require complex training strategies for streaming scenarios \cite{las1,las2}. RNNT \cite{rnnt} provides a natural alternative in streaming scenarios but has high training complexity and inference latency rendering it difficult to use in a real-world setting without complex optimizations/modifications \cite{rnnt1, rnnt2}.

There have been many attempts to improve the accuracy of CTC models that preserve their training and inference simplicity. \citet{hctc} leverages hierarchical structure in the speech by adding auxiliary losses to train a CTC-based acoustic-to-subword model. Their hierarchical CTC (HCTC) model predicts different text segmentations in a fine-to-coarse fashion. Recent studies have explored the use of attention in CTC models to implicitly relax the conditional independence assumption by enriching the features using other time frames. \citet{ctc_attention1} uses component attention and implicit language model to enrich the context while \citet{ctc_attention2} evaluates a fully self-attention-based network with CTC. In this work, we explore how augmenting an LSTM-based network with windowed self-attention can help improve the transcription while preserving streaming capability.

Another line of work in improving the output of streaming models is the second pass rescoring that uses an additional (usually non-streaming) component to re-rank the streaming model's hypotheses \cite{two_pass}. While we also rescore the candidate hypotheses at the last step, our system doesn't employ any external acoustic model to do so and leverages the hierarchical losses that are part of the model itself.

For addressing EOS detection, conventional approaches use VAD models with a threshold on silence amount. This may lead to early termination of user speech. \citet{eoq} addresses this by training an EOQ (End-of-Query) classifier which performs better than VAD but is still optimized independent of the ASR system. VAD based on output CTC labels has also been explored to detect EOS based on the length of non-speech (blank) region \cite{ctc_eos}. \citet{rnnt_eos} jointly train an RNNT model for EOS detection by using and extra $</s>$ token with early and late penalties. Prediction of $</s>$ token by the model during inference marks as the signal for EOS. We follow a similar approach where we train the model with early and late penalties. During inference, we use a dynamic threshold on $</s>$ probability to detect the endpoint before decoding the text.

\section{Methodology}
\label{sec:method}

\subsection{Model Architecture}
\label{ssec:model}

\begin{figure}[htbp]
  \centering
  \includesvg[width=0.9\columnwidth]{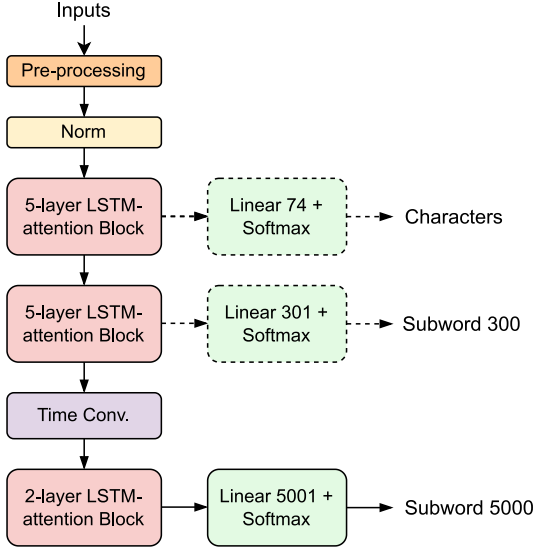}
  \caption{ASR model. Characters, Subword 300 and Subword 5000 are used as targets to compute the CTC losses at resp. levels.}
  \label{fig:asr_model}
\end{figure}

Inspired by~\citet{hctc}, we build a 3-level HCTC architecture based on LSTM and attention as shown in Fig.~\ref{fig:asr_model}. Going in a fine-to-course fashion, the model predicts characters (73 tokens), short subwords (300 tokens) and long subwords (5000 tokens) at the respective levels. Each level consists of an N-layer LSTM-attention block (N being 5, 5 and 2) followed by a linear softmax layer. A time convolution layer with a kernel size of 5 and a stride of 3 after the second level reduces the number of time steps to one-third. This helps emit longer subwords at the third level by increasing the context and receptive field of a time frame. Along with the HCTC loss, we use label smoothing~\cite{lab_smooth} by adding a negative entropy term to it. This mitigates overconfidence in output distributions leading to improved transcription. Mathematically, the loss for a given training sample, $(x,y) = (x,\{y_{char},y_{s300},y_{s5k}\})$, is:
\begin{align*}
L(x,y) &= \sum_k\Big[CTCLoss(x,y_k)\\ 
&\qquad\quad-\lambda\sum_tEntropy(P_k(:|x_t))\Big]\\
&= \sum_k\Big[-log(P(y_k|x))\\ 
&\qquad\quad+\lambda\sum_{t,v}P_k(v|x)log(P_k(v|x))\Big]\\
\end{align*}

\begin{figure}[htbp]
  \centering
  \includesvg[height=1\columnwidth]{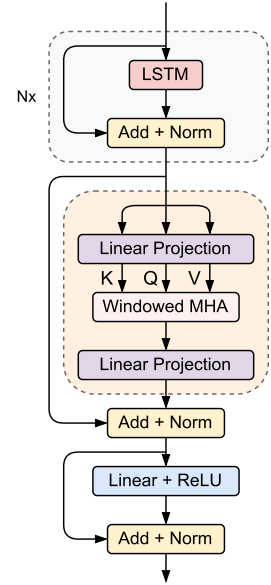}
  \caption{N layer LSTM-attention Block}
  \label{fig:lstm_block}
\end{figure}

For an N-layer LSTM-attention block (Fig.~\ref{fig:lstm_block}), we stack N LSTM layers with 700 hidden dimensions which are followed by a dot-product based multi-headed self-attention layer (MHA)~\cite{transformer}. We use 8 attention heads and project the input to 64-dimensional key, query and value vectors for each head. We project back the 512 (8x64) dimensional output to 700 dimensions and pass it through a linear layer with ReLU activation. To retain the model's streaming capabilities, we restrict the attention to a 5-frame window (t$\pm$2) instead of complete input i.e., for input features $f_t$, we use $Q(f_t)$ as the query vector and $K(f_{t-2:t+2})$, $V(f_{t-2:t+2})$ as key-value vectors where $Q$,$K$ and $V$ are linear projections. To improve the gradient flow, we add a skip connection and layer normalization after each layer.

We use 80 filterbanks from standard log-mel-spectrogram as inputs, computed with a window of 20ms, a stride of 10ms, and an FFT size of 512. To prevent overfitting, we use time-frequency masking \cite{specaugment} during training. We also stack five adjacent frames with a stride of three, giving an input feature vector of 400 dimensions with a receptive field of 60ms and stride of 30ms for each time step. Windowed MHA and time convolution increase overall receptive field and stride to 780ms and 90ms resp. Consequently, our model has a forward lookahead of 390ms when deployed in a streaming mode.

\subsection{Speech End-pointing}
\label{ssec:speech_endpoint}

Once we have a trained ASR model, we augment the vocabulary with an additional $</s>$ token and use forced alignment to get the ground truth speech endpoints. We use the output from 1st (character) level of the ASR model for alignment as it has the least lookahead and empirically works better than the output from other blocks. We append the extra $</s>$ token at the end of each transcript and add early-late (EL) penalties \cite{rnnt_eos} to the training loss to fine-tune the model for a few more iterations. EL penalties penalize the model for predicting $</s>$ too early or too late. During online inference, we determine if the current time step ($t$) is the speech endpoint by evaluating the following conditions:
\begin{itemize}
    \item There is at least one word in output text - to avoid termination before the user starts speaking;
    \item $</s>$ is the most probable token among all vocab items i.e., $P_t(</s>) \geq P_t(:)$ - call this an EOS peak;
    \item $P_t(</s>) \geq threshold_t = \alpha^{1+n_t/\beta}$ where $n_t$ is the number of EOS peaks before time $t$.
\end{itemize}
Thus, the earliest time step satisfying the above conditions is the EOS. Here $\alpha$ controls the aggressiveness of EOS detection as decreasing $\alpha$ decreases the EOS threshold for all time steps resulting in an earlier EOS signal. Empirically, we observe that the model gives a lower probability to $</s>$ token after each EOS peak. To address this, we add an $n_t/\beta$ term that gradually reduces the threshold whenever an EOS peak appears, giving an additional (but marginal) reduction in latency. For audios where the above conditions are never satisfied, a combination of a small independent VAD model and a maximum time limit works as a backup.


\begin{table*}[htbp]
  \centering
  \begin{tabular*}{0.9\textwidth}{l @{\extracolsep{\fill}} rrrrrr}
  \hline
    \textbf{Model} & \multicolumn{3}{c}{\textbf{\%WER}} & \multicolumn{3}{c}{\textbf{Mean EOS Latency}} \\
    & \textbf{all} & \textbf{clean} & \textbf{noisy} & \textbf{all} & \textbf{clean} & \textbf{noisy} \\
    \hline
    Google Speech-to-Text API* & 13.14 & 12.62 & 16.30 \\
    LSTM-attention HCTC (all data)              & 4.03 & 3.11 & 9.52 & 2858 & 2457 & 5080 \\
    + fine-tune on target domain & 3.75 & 3.03 & 8.02 & 2858 & 2457 & 5080 \\
    + EL penalty (our best model) & \textbf{3.69} & \textbf{2.95} & \textbf{8.12} & 2858 & 2457 & 5080 \\
    + EOS detection (without $n/\beta$ term) & 4.77 & 4.10 & 8.75 & 1565 & 1268 & 3215 \\
    + EOS detection (with $n/\beta$ term) & \textbf{4.78} & \textbf{4.19} & \textbf{8.32} & \textbf{1525} & \textbf{1242} & \textbf{3096} \\
    \hline
    Reduction in Latency         & & & & 1333 & 1215 & 1985 \\
  \hline
  \end{tabular*}
  
  \caption{Results for the best model with and without EOS detection. EOS detection reduces mean latency by $\sim$1300 ms. *Google's API has a much higher WER because it is trained for open domain whereas our data is in e-commerce domain and also has background noise.}
  \label{tab:main_results}
\end{table*}

\subsection{Decoding and Re-scoring}
\label{ssec:decoding}

For each chunk of input audio stream, we use prefix beam search, with a beam size of 1000 hypotheses, to decode the text from probability distribution given by the last (subword 5000) level. We use the same probability distribution to detect EOS as well. When we observe an EOS or the stream ends, a 5-gram KenLM and HCTC loss (sum of CTC losses from all levels) are used to re-rank and select the best hypothesis from the top 100 candidates. We use grid search to find the weights of the scores.

\section{Dataset and Training Setup}
\label{sec:setup}

Queries from E-commerce Voice Search are our primary source of data. We also collect speech from other sources like on-call customer support, crowdsourced read-speech, etc., to augment training data. We transcribe all the utterances, except read-speech, using an existing ASR system and manually correct them. The ASR system that generates reference transcripts progressively improves as part of model iterations. Collectively, the training datasets amount to $\sim$14M audio-text pairs (8M from the target domain and 6M from other) or roughly 22.5k hours of audio. For evaluation, we randomly sample $\sim$19k audios from e-commerce voice search queries, transcribe it manually (without any reference text) and reduce the human error by using multiple iterations of verification.

We categorize the test set into clean and noisy subsets, containing $\sim$16k and $\sim$3k samples resp. Clean utterances are audios where only one speaker's speech is intelligible.  Noisy utterances are those where more than one speaker has intelligible speech (overlapping or non-overlapping). In noisy utterances, the primary speaker is the user whose utterance is more relevant for the e-commerce voice search application. Note that clean utterances may also have non-intelligible secondary speakers. We train and evaluate the model to transcribe only the primary speaker's speech while ignoring the rest.

For training KenLM and Sentencepiece models, we use a large corpus comprising text from various sources like transcribed voice search queries and on-call customer support queries, customer support chatbot queries, and product catalogues.

We use a cyclical learning rate (LR) \cite{cyclic_lr} with Adam optimizer to train the ASR model for 200k iterations with a batch size of $\sim$55 minutes. Training the model on two A100 (40 GB) GPUs takes $\sim$50 hours. For EOS detection, we fine-tune the model with EL penalties for an additional 48k iterations ($\sim$12 hours).

\section{Results and Discussion}
\label{sec:results}

\begin{figure}[htbp]
  \centering
  \includesvg[width=0.9\columnwidth]{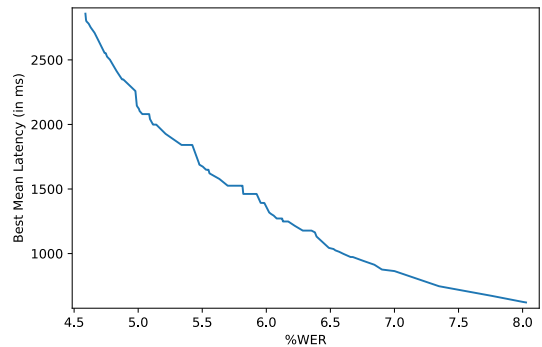}
  \caption{Mean EOS latency vs \%WER as we change $\alpha$ and $\beta$.}
  \label{fig:wer_vs_latency}
\end{figure}

We report WER and mean EOS latency on the test set for evaluating the performance of our model in Table~\ref{tab:main_results}. We get the best results when the model is first pre-trained on all the data and then fine-tuned on the target domain, followed by fine-tuning with EL penalties. To see how $\alpha$ and $\beta$ affect the results, we do a sweep over both the parameters and plot mean EOS latency vs \%WER in Fig.~\ref{fig:wer_vs_latency}. For further analysis, we consider the point with $\alpha=0.8$ and $\beta=2.0$ that gives us a WER of 4.78\% and a reduction of 1333 ms in mean EOS latency with EOS coverage (fraction of audios receiving an EOS signal) of 64.13\%. Our model performs significantly better than Google Speech-to-Text API, which is expected since Google's API is trained for the open domain, but our data is in the e-commerce domain. The evaluation utterances also have a lot of noise which our model is more robust to as it is trained on similar data.

\begin{table}[htbp]
  \centering
  \begin{tabular}{lrrr}
  \hline
    \textbf{Model} & \textbf{\%WER} & \textbf{$\Delta$WER} \\ \hline
    LSTM-attention HCTC & 5.37 & \multicolumn{1}{l}{} \\
    - Windowed MHA & 5.94 & 9.60\% \\
    - HCTC rescoring & 6.19 & 4.04\% \\
    - HCTC loss & 6.62 & 6.50\% \\
    - Skip connections & 7.68 & 13.80\% \\
    \quad (= Baseline LSTM CTC) & & \\
  \hline
  \end{tabular}
  \caption{Change in WER when each component is removed. All results are with LM rescoring using the same KenLM.}
  \label{tab:ablation_results}
\end{table}

\begin{table*}[htbp]
\centering
\begin{tabular}{lll}
\hline
\textbf{Ground truth} & \textbf{ASR output} & \textbf{Reason for error} \\ \hline
\textbf{\underline{no search impact}} \\
\textbf{\color{orange} mixer} machine & \textbf{\color{red} mixture} machine & wrong pronunciation \\
ooni kapda & \textbf{\color{red} baby} ooni kapda & multiple speakers with similar voice \\
\textbf{\color{orange} sasta} sasta mobile vivo ka & sasta mobile vivo ka & overlapping speakers \\
choli photos \textbf{\color{orange} choli photos} & choli photos & repetition after EOS \\
chappal \textbf{\color{orange} slipper} & chappal & repetition after EOS (in other lang.) \\ \hline
\textbf{\underline{-ve search impact}} \\
\textbf{\color{orange} great} cycle & \textbf{\color{red} grey} cycle & background noise \\
\textbf{\color{orange} capacitor} & \textbf{\color{red} cap sitter} & wrong pronunciation \\
earing & \textbf{\color{red} car} earing & multiple speakers with similar voice \\
\textbf{\color{orange} atlas} three chaubis \textbf{\color{orange} inch} & \textbf{\color{red} headlight} three chaubis \textbf{\color{red} pin} & overlapping speakers \\
\textbf{\color{orange} joota} & \textbf{\color{red} guitar} & two eligible primary speakers \\
oppo a thirty three \textbf{\color{orange} back cover} & oppo a thirty three & additional information after EOS \\
\hline
\end{tabular}
\caption{Examples of different types of errors. The upper section of the table shows examples where the mistakes don't have any search impact, and the lower section shows the ones having a negative effect. Red indicates incorrect words and insertion errors, and orange indicates deletions and correct counterparts of erroneous words.}
\label{tab:examples}
\end{table*}

To understand how modifications in the architecture contribute to improving the accuracy of the vanilla LSTM CTC model, we conduct an ablation study and report the WER in Table~\ref{tab:ablation_results}. We train these models for 200k iterations on a reduced dataset of $\sim$5500 hours sampled from the target domain. As seen from the table, windowed MHA improves the WER by 9.6\%. Intuitively, the improvement comes from an increased receptive field (780ms with vs 180ms without attention) and the ability to extract better context from neighbouring frames using self-attention. HCTC loss forces the model to learn hierarchical structure in the speech at multiple levels - from characters to short subwords and then long subwords. The model can then utilize this structure to achieve more accurate predictions. Adding auxiliary losses at intermediate levels helps the convergence as well. The hierarchical loss also facilitates the rescoring since the combination of losses acts like an ensemble of ranking models. Together, HCTC loss and rescoring give a relative improvement of 10.28\%. Finally, skip connections improve the gradient flow in training, which further helps the convergence, improving the WER by 13.80\%. These modifications, when combined, result in a significant total relative improvement of $\sim$30\% in WER over the baseline.

\begin{table}[htbp]
  \centering
  \begin{tabular}{lrrr}
  \hline
  \textbf{Model} & \multicolumn{3}{c}{\textbf{\%WER}} \\
  & \textbf{all} & \textbf{clean} & \textbf{noisy} \\
  \hline
    \textbf{\underline{Streaming models}} \\
    LSTM CTC & 7.68 & 6.50 & 14.70 \\
    LSTM-atten. HCTC & 5.37 & 4.57 & 10.12 \\
    Str. Conformer CTC & 5.30 & 4.33 & 11.01 \\
  \hline
    \textbf{\underline{Non-streaming models}} \\
    BiLSTM CTC & 5.37 & 4.64 & 9.68 \\
    BiLSTM-atten. HCTC & 4.65 & 3.97 & 8.69 \\
    Transf. AED+CTC & 4.37 & 3.30 & 10.72 \\
    \cite{transformer_ctc} &&&\\
  \hline
  \end{tabular}
  \caption{Our model in comparison with others (details in Sec.~\ref{ssec:comparison}). All models are made similar in size and trained on $\sim$5500 hours.}
  \label{tab:comparison}
\end{table}



\subsection{Comparison with other models}
\label{ssec:comparison}

In addition to the baseline LSTM CTC (Table~\ref{tab:ablation_results}), we also compare our model with a non-streaming BiLSTM version, and a streaming Conformer CTC inspired by \cite{conformer_rnnt}. For Conformer CTC, we use the causal encoder-only network and train it using CTC loss. As evident from the results in Table~\ref{tab:comparison}, the discussed modifications help bridge the gap between streaming LSTM and non-streaming BiLSTM CTC models. The streaming Conformer CTC also performs only marginally better than our LSTM-attention HCTC model while it has much higher training complexity and inference latency.

We evaluate a bidirectional version of our model to analyse the consistency of these improvements. Observe that the same modifications improve the BiLSTM CTC model by a relative 13.4\%, vs 30\% the LSTM CTC model because BiLSTM already has access to full future context, limiting the scope of improvement. Even then, it performs significantly better than a vanilla BiLSTM CTC model and only slightly worse than a Transformer AED+CTC model \cite{transformer_ctc}. Thus, these modifications also reduce the gap between LSTM and transformer-based ASR models for voice search in both - streaming and non-streaming settings. One explanation could be that transformers usually have an advantage in capturing long-term dependencies. This doesn't help as much for speech recognition on short utterances as in our dataset, where audios usually are 4-6 seconds long with an average of 3.34 spoken words. For a fair comparison, we ensure all models are similar in size and use the same KenLM for rescoring.

\subsection{Error Analysis}
\label{ssec:analysis}

To understand the errors better, we analyze 50 random utterances each from clean and noisy subsets where the model makes mistakes. The most common reasons for errors in the clean subset are - wrong pronunciation and background noise. For noisy utterances, multiple speakers with a similar voice, overlapping speakers, and more than one eligible primary speakers contribute to additional errors. Table~\ref{tab:examples} lists some examples demonstrating these reasons. We also observe that around 62\% of the mistakes in the evaluation set have no negative impact on search. In these cases, the errors are usually in stop words or produce a variant of the reference word which can be used, like singular vs plural or the same word with a different spelling.

When using EOS detection, there are additional errors due to early termination in 2.24\% of the utterances. In all such cases, EOS is detected prematurely because of a pause in the speech. Usually, after this pause, the user repeats their query, adds more information, or corrects it. In around 47\% of the cases, not capturing this additional speech has no negative impact on search. In the rest 53\% cases, i.e. 1.19\% of the total samples, the missed utterance usually has more information about the query, added by the user, that could have helped in refining the search results.

\subsection{Conclusions}
\label{ssec:conclusions}

This work focuses on developing a robust and efficient streaming ASR model for Hinglish Voice Search. We achieve this by utilizing an LSTM-attention architecture and employing the HCTC loss. We explore architectural modifications that help bridge the accuracy gap between streaming and non-streaming LSTM-based ASR models.

Our proposed model performs on par with a streaming conformer-based system but offers the advantage of lower latency. Additionally, we present a straightforward method to integrate End-of-Speech (EOS) detection with CTC-based models, requiring only a small number of additional training iterations and utilizing simple thresholding during inference.

The simplicity and low latency of our model contribute to a fast and accurate voice search experience, making it an appealing solution for practical applications.

\section*{Limitations and Future Work}

In our study, we focused on a high-resource setting with access to approximately 22.5k hours of labeled speech data. While we compared our models with conformer and transformer-based AED and CTC models, we did not include RNNT models due to their higher compute resource requirements. To accommodate deployment constraints, we employed a smaller model with approximately 60 million parameters, which limited its performance.

Moving forward, our future work aims to explore the potential benefits of leveraging large unsupervised datasets and larger models to further enhance our system and extend its applicability to other Indian languages, which typically have less available data compared to Hinglish. Building upon our previous success in adapting a non-streaming model for end-to-end speech-to-intent detection in customer support voicebots \cite{e2e_s2i}, we are motivated to investigate the feasibility of developing a single joint model for Automatic Speech Recognition (ASR), End-of-Speech (EOS) detection, and Spoken Language Understanding (SLU). Additionally, we are keen on exploring the development of multilingual ASR models.

\bibliography{anthology,custom}
\bibliographystyle{acl_natbib}

\end{document}